\documentstyle{mn}

\newif\ifAMStwofonts
\ifoldfss
  \ifCUPmtlplainloaded \else
    \NewTextAlphabet{textbfit} {cmbxti10} {}
    \NewTextAlphabet{textbfss} {cmssbx10} {}
    \NewMathAlphabet{mathbfit} {cmbxti10} {} % for math mode
    \NewMathAlphabet{mathbfss} {cmssbx10} {} %  "   "    "
  \fi
  \ifAMStwofonts
    \ifCUPmtlplainloaded \else
      \NewSymbolFont{upmath} {eurm10}
      \NewSymbolFont{AMSa} {msam10}
      \NewMathSymbol{\upi}     {0}{upmath}{19}
      \NewMathSymbol{\umu}     {0}{upmath}{16}
      \NewMathSymbol{\upartial}{0}{upmath}{40}
      \NewMathSymbol{\leqslant}{3}{AMSa}{36}
      \NewMathSymbol{\geqslant}{3}{AMSa}{3E}

      \let\leq=\leqslant \let\le=\leqslant
       \let\ge=\geqslant
    \fi
  \fi
\fi % End of OFSS

\ifnfssone
  \newmathalphabet{\mathit}
  \addtoversion{normal}{\mathit}{cmr}{m}{it}
  \addtoversion{bold}{\mathit}{cmr}{bx}{it}
  \newmathalphabet{\mathbfit} % math mode version of \textbfit{..}
  \addtoversion{normal}{\mathbfit}{cmr}{bx}{it}
  \addtoversion{bold}{\mathbfit}{cmr}{bx}{it}
  \newmathalphabet{\mathbfss} % math mode version of \textbfss{..}
  \addtoversion{normal}{\mathbfss}{cmss}{bx}{n}
  \addtoversion{bold}{\mathbfss}{cmss}{bx}{n}
  \ifAMStwofonts
    \ifCUPmtlplainloaded \else
      \UseAMStwoboldmath
      \makeatletter
      \new@mathgroup\upmath@group
      \define@mathgroup\mv@normal\upmath@group{eur}{m}{n}
      \define@mathgroup\mv@bold\upmath@group{eur}{b}{n}
      \edef\UPM{\hexnumber\upmath@group}
      \new@mathgroup\amsa@group
      \define@mathgroup\mv@normal\amsa@group{msa}{m}{n}
      \define@mathgroup\mv@bold\amsa@group{msa}{m}{n}
      \edef\AMSa{\hexnumber\amsa@group}
      \makeatother
      \mathchardef\upi="0\UPM19
      \mathchardef\umu="0\UPM16
      \mathchardef\upartial="0\UPM40
      \mathchardef\leqslant="3\AMSa36
      \mathchardef\geqslant="3\AMSa3E

      \let\leq=\leqslant \let\le=\leqslant
       \let\ge=\geqslant
    \fi
  \fi
\fi % End of NFSS release 1

\ifnfsstwo
  \DeclareMathAlphabet{\mathbfit}{OT1}{cmr}{bx}{it}
  \SetMathAlphabet\mathbfit{bold}{OT1}{cmr}{bx}{it}
  \DeclareMathAlphabet{\mathbfss}{OT1}{cmss}{bx}{n}
  \SetMathAlphabet\mathbfss{bold}{OT1}{cmss}{bx}{n}
  \ifAMStwofonts
    \ifCUPmtlplainloaded \else
      \DeclareSymbolFont{UPM}{U}{eur}{m}{n}
      \SetSymbolFont{UPM}{bold}{U}{eur}{b}{n}
      \DeclareSymbolFont{AMSa}{U}{msa}{m}{n}
      \DeclareMathSymbol{\upi}{0}{UPM}{"19}
      \DeclareMathSymbol{\umu}{0}{UPM}{"16}
      \DeclareMathSymbol{\upartial}{0}{UPM}{"40}
      \DeclareMathSymbol{\leqslant}{3}{AMSa}{"36}
      \DeclareMathSymbol{\geqslant}{3}{AMSa}{"3E}

      \let\leq=\leqslant \let\le=\leqslant
       \let\ge=\geqslant
    \fi
  \fi
\fi % End of NFSS release 2

\ifCUPmtlplainloaded \else
  \ifAMStwofonts \else % If no AMS fonts
    \def\upi{\pi}
    \def\umu{\mu}
    \def\upartial{\partial}
  \fi
\fi
\title{Testing the homogeneous synchrotron self Compton model 
for gamma ray production in Mrk 421}
\author[W. Bednarek and R.J. Protheroe]
       {W. Bednarek$^*$ and R.J. Protheroe \\
Department of Physics and Mathematical Physics,
The University of Adelaide, Adelaide, Australia 5005.\\
$^*$Permanent address: University of \L\'od\'z, 90-236\L\'od\'z, 
ul. Pomorska 149/153, Poland.  
               }
\date{University of Adelaide preprint ADP-AT-97-3, submitted to MNRAS}

\pubyear{199}

\begin{document}

\maketitle

\label{firstpage}

\begin{abstract}

Based on the detected variability time scales of X-ray and TeV
gamma-ray emission, and the observed multiwavelength photon
spectrum, of Mrk 421 we place constraints on the allowed
parameter space (magnetic field and Doppler factor of the
emission region) for the homogeneous synchrotron self-Compton
model. The spectra calculated for the allowed parameters are
marginally consistent with the available spectral information
above $\sim 1$ TeV reported by the Whipple Observatory in the
case of a 1 day flare time scale. However, for the recently
reported very short duration flares varying on a time scale of 15
min, the calculated spectra are significantly steeper, suggesting
that the homogeneous synchrotron self Compton model has problems
in describing the relatively flat observed spectra extending
above a few TeV.  We determine the maximum ratio of TeV gamma-ray
luminosity to X-ray luminosity during flaring which is allowed by
the homogeneous synchrotron self-Compton model for the case of no
significant photon-photon absorption in the source.

\end{abstract}
 

\section{Introduction}

Very high energy (VHE) $\gamma$-ray emission has been detected in
recent years from two BL Lac objects Mrk 421 (Punch et
al.~1992, Petry et al.~1996) and Mrk 501 (Quinn
et al.~1996). The emission varies significantly on
different time scales, from weeks and days (Kerrick et
al.~1995, Schubnell et al.~1996, Buckley et
al.~1996), to fractions of an hour (Gaidos et
al.~1996).  Multiwavelength observations of Mrk 421 show
that the TeV $\gamma$-ray flares are simultaneous with the X-ray
flares observed by the ASCA satellite, and that the power emitted
in these two energy ranges is comparable (Takahashi et
al.~1996, Macomb et al.~1995, Buckley et
al.~1996). However during this same period, the amplitude
of variations in the lower energy emission (UV--optical) and the
$\gamma$-ray emission in the EGRET energy range was much less
than that of the very high energy $\gamma$-ray emission (Macomb
et al.~1995, Buckley et al.~1996, Lin et
al.~1994).  VHE emission from Mrk 421 has been observed
up to $\sim 4$ TeV in the quiescent state (Mohanty et
al.~1993), and up to at least $\sim 8$ TeV in the high
state (Krennrich et al.~1997). Emission extending to
similar energies has also been observed recently from Mrk 501
(Breslin et al.~1997).  Summing the HEGRA data from
several nearby blazars (including Mrk 421) there is some evidence
($6.5 \sigma$) of $\gamma$-ray emission above 50 TeV (Meyer \&
Westerhoff~1996, Meyer~1997), and for Mrk 421 alone
the significance is $3.8\sigma$.

Gamma ray emission from active galactic nuclei (AGN) is often
interpreted in terms of the homogeneous ``synchrotron
self-Compton model'' (SSC) in which the low energy emission (from
radio to X-rays) is synchrotron radiation produced by electrons
which also up-scatter these low energy photons into high energy
$\gamma$-rays by inverse Compton scattering (ICS) (Macomb et
al.~1995, Inoue \& Takahara~1996, Bloom \& Marscher~1996,
Mastichiadis \& Kirk~1997). In this model all the radiation comes
from this same region in the jet. Such a picture can naturally
explain synchronized variability at different photon
energies. More complicated (inhomogeneous) SSC models are also
proposed which postulate that the radiation at different energies
is produced in different regions of the jet (e.g.  Ghisellini et
al.~1985, Maraschi et al.~1992). It has also been argued that the
$\gamma$-ray emission from Mrk 421 can be explained by electrons
scattering in the Klein-Nishina regime (Zdziarski \&
Krolik~1993).

Photon-photon pair production on the infrared background
radiation was expected to prevent observation above $\sim 1$ TeV.
The observation of $\gamma$-rays from Mrk~421 up to 8~TeV without
a cut-off in the spectrum, and certainly up to 50~TeV, was
unexpected based on calculations for reasonable models of the
infrared radiation field (see e.g. Stecker and De Jager 1997 and
references therein).  Even if cascading is included in the
infrared background (Protheroe and Stanev 1993, Entel and
Protheroe 1995) observation of $\gamma$-rays of these energies
seemed unlikely.  However, the infrared background is not well
known and it may be possible to observe the nearest blazars up to
energies somewhat below $\sim 100$ TeV where absorption on the
cosmic microwave background will give a sharp cut-off.  In our
calculations below, we shall therefore neglect photon-photon pair
production on the infrared background.

The purpose of this paper is to confront the homogeneous SSC model
with the results of recent observations and, if possible, to
derive the parameters of the emission region in the jet from
which this radiation originates, i.e. its Doppler factor and
magnetic field strength.

\section{Constraints on a homogeneous SSC model}

Let us consider relativistic electrons confined in a ``blob''
which moves along the jet with the Doppler factor $D$ and has
magnetic field $B$.  In the homogeneous SSC model the radii of
the emission regions of low energy photons ($r_l$), X-ray photons
($r_X$), and TeV $\gamma$-rays ($r_{\gamma}$) are the
same. This region is constrained by the variability time scale
observed, e.g. in TeV $\gamma$-rays, $t_{\rm var}$ (s),
\begin{eqnarray}
r_l = r_{\gamma} = r_X \approx 0.5 c D t_{\rm var}.
\label{eq1}
\end{eqnarray}  
\noindent
The differential photon density in the blob frame of low energy
synchrotron photons (phot. MeV$^{-1}$ cm$^{-3}$) is then given by
\begin{eqnarray}
n(\epsilon') \approx {{4d^2 F(\epsilon)}\over{c^3 t_{\rm var}^2 D^4}},
\label{eq2}
\end{eqnarray}
\noindent
where $d\approx 187$ Mpc is the distance to Mrk 421 (for $H_0 =
50$ km s$^{-1}$ Mpc$^{-1}$, and $z = 0.031$), $\epsilon =
D\epsilon'$ and $\epsilon'$ are the photon energies in the
observer's and the blob rest frames, and $c$ is the velocity of
light.  

The differential photon flux in the optical to X-ray region
observed from Mrk 421 during the 16 May 1994 flare (photons
cm$^{-2}$ s$^{-1}$ MeV$^{-1}$) can be approximated by a broken
power-law,
\begin{eqnarray}
F(\epsilon) \approx \cases {b_1 \epsilon^{-\beta_1} & $
10^{-3}\epsilon_{b} < \epsilon \le \epsilon_{b}$,\cr b_2
\epsilon^{-\beta_2} & $\epsilon \ge \epsilon_{b}$,\cr}
\label{eq3}
\end{eqnarray}
where $\beta_1 = 1.8$, $\beta_2 = 2.3$, $b_1 = 6.9\times 10^{-4}$
and $b_2 = 2.8\times 10^{-5}$ are obtained from ASCA
observations, and $\epsilon_{b} = 1.65\times 10^{-3}$ MeV is the
energy at which a break in the synchrotron spectrum is observed.
Eq.~\ref{eq3} is based on the peak 2--10 keV luminosity and
spectral index given in Fig.~3 of Takahashi et al. (1996) (for
$\epsilon > \epsilon_{b}$), and from Fig.~2 of Macomb et
al. (1995) we estimate the spectrum for $\epsilon \le
\epsilon_{b}$.  Note that for at least 2 decades below
$10^{-3}\epsilon_{b}$ the spectrum is uncertain.

The shape of synchrotron spectrum defines the shape of the electron 
spectrum in the blob rest frame which can be approximated by 
\begin{eqnarray}
{{dN}\over{d\gamma'}} \approx \cases {a_1 \gamma'^{-\alpha_1} &
$0.032\gamma'_{b} < \gamma' \le \gamma'_{b}$,\cr a_2
\gamma'^{-\alpha_2} & $\gamma' \ge \gamma'_{b}$,\cr}
\label{eq4}
\end{eqnarray}
\noindent
where $\gamma'$ is the Lorentz factor in the blob frame,
\begin{eqnarray}
\gamma'_{b} = (2\epsilon_{b}/D\epsilon_B)^{1/2},
\label{eq4a}
\end{eqnarray}
\noindent
$\alpha_1=2.6$, $\alpha_2=3.6$, $\epsilon_B = m_e c^2 B/B_{cr}$,
$B_{cr}=4.414\times 10^{13}$ G, $m_e$ is the electron rest mass,
$a_1 = a_2/\gamma'_{b}$, and $a_2$ can be obtained from fitting
the observations.  Note that below
$10^{-3/2}\gamma'_{b}=0.032\gamma'_{b}$ the spectrum is uncertain.

The spectrum of Mrk 421 shows two clear bumps which, during the
outburst stage, extend up to at least $\sim 10$ keV (Takahashi et
al.~1996), and $\sim 8$ TeV (Krennrich et al.~1997).  These
multiwavelength observations of Mrk 421 allow us to define the
ratio $\eta$ of the power emitted at a $\gamma$-ray energy,
$E_\gamma$, at which the emission is due to Compton scattering,
to the power emitted at an energy, $\epsilon$, at which the
emission is due to X-ray synchrotron radiation,
\begin{eqnarray}
\eta = ({{dN}\over{dE_\gamma dt}} E_\gamma^2)/({{dN}
\over{d\epsilon dt}} \epsilon^2) =
({{dN}\over{dE'_\gamma dt'}} {E'}_{\gamma}^{2})/
({{dN}\over{d\epsilon' dt'}}{\epsilon'}^{2}), 
\label{eq5}
\end{eqnarray}
\noindent
where the primed quantities are measured in the blob frame. For
the power at $\gamma$-ray energies we adopt the value reported
for the threshold of the Whipple telescope at $E_\gamma = 0.3$
TeV (Macomb et al.~1996), and for the power at X-ray synchrotron
energies we take the value corresponding to the peak emission at
$\epsilon=\epsilon_b$ (Takahashi et al.~1996). For these two
energies $\eta\approx 1.2$.

The synchrotron spectrum at $\epsilon'$ in the above formula 
(Eq.~\ref{eq5}) can be obtained approximately analytically from the 
relation
\begin{eqnarray}
\epsilon' {{dN}\over{d\epsilon' dt}} d\epsilon' \approx 
{{dN}\over{d\gamma'}} d\gamma' b_{\rm syn}(\gamma'),
\label{eq6}
\end{eqnarray}
\noindent
where $dN/d\gamma'$ is the electron spectrum (Eq.~\ref{eq4}). 
The characteristic energy of synchrotron photons is given by
\begin{eqnarray}
\epsilon' \approx 0.5\epsilon_B\gamma'^2, 
\label{eq7}
\end{eqnarray}
\noindent
the energy loss rate of electrons is $b_{\rm syn}(\gamma') = k U_B
\gamma'^2$, where $k = 4 c \sigma_T/3$, $\sigma_T$ is the Thomson
cross section, and $U_B\approx 2.5\times 10^4 B^2$ (MeV
cm$^{-3}$) is the magnetic field energy density. The synchrotron
spectrum emitted by electrons with power-law spectral index
$\alpha$, multiplied by the square of the photon energy, is given
by
\begin{eqnarray}
{{dN}\over{d\epsilon' dt'}}{\epsilon'}^{2} \approx
{{2a k U_B{\epsilon'}^{2}}\over{\epsilon_B^2}} 
\left({{2\epsilon'}\over{\epsilon_B}}\right)^{-(\alpha+1)/2}.
\label{eq8}
\end{eqnarray} 
The ICS part of the Eq.~\ref{eq5} cannot be obtained analytically
in the general case because of the complicated form of the
Klein-Nishina cross section, and so we compute this numerically
using
\begin{eqnarray}
{{dN}\over{dE'_\gamma dt'}}{E'}_{\gamma}^{2} = 
{E'}_{\gamma}^{2} \int_{\gamma'_{\rm min}}^{\infty}  
{{dN}\over{d\gamma'}}\int_{\epsilon'_{\rm min}}^{\infty} {{dN(\gamma', 
E'_\gamma)}\over{dt' d\epsilon' dE'_\gamma}} d\epsilon' d\gamma',
\label{eq20}
\end{eqnarray} 
\noindent
where $\gamma'_{\rm min}\approx E'_\gamma/m_e c^2$,
$\epsilon'_{\rm min} =E'_\gamma/[4\gamma' (\gamma' -
E'_\gamma/m_e c^2)]$, $E'_\gamma = E_\gamma/D$, and $dN(\gamma',
E'_\gamma)/dt' d\epsilon' dE'_\gamma$ is the ICS spectrum (see
Eq.~2.48, in Blumenthal \& Gould~1970) produced by electrons with
Lorentz factor $\gamma'$ which scatter synchrotron photons in the
blob having the spectrum given by Eqs.~\ref{eq2} and~\ref{eq3}.

Having determined the spectra in Eq.~\ref{eq5}, we can now
investigate the parameter space (magnetic field strength in the
blob, $B$, and Doppler factor, $D$) for the homogeneous SSC model
which is consistent with the value of $\eta = 1.2$. In
Figs.~\ref{fig1}(a) and \ref{fig1}(b) we show the allowed value
of $B$ as a function of $D$ (thick full curves) for the case of
outbursts as reported by the Whipple Observatory which varied on
(a) a $\sim 1$ day (Buckley et al.~1996, Schubnell et al.~1996)
and (b) a $\sim 15$ min (Gaidos et al.~1996) time scale.

\subsection{Limits from variability time scales}

The variability time scales observed in TeV $\gamma$-rays by the
Whipple Observatory, and the reports of simultaneous flares
observed in X-rays by ASCA allow us to place a further constraint
on the homogeneous SSC model. A significant decrease in the
observed TeV $\gamma$-ray and X-ray fluxes may only occur if the
electrons have sufficient time to cool during the flare,
\begin{eqnarray}
{t'}_{\rm cool} \leq t_{\rm var} D.
\label{eq13}
\end{eqnarray}
\noindent
The cooling time scale for synchrotron losses of electrons with Lorentz 
factor $\gamma'_{b}$, which contribute mainly to synchrotron photons at 
the peak of the spectrum, is given by
\begin{eqnarray}
{t'}_{\rm cool}^{\rm syn} = {{m_e c^2}\over{kU_B\gamma'_{b}}}.  
\label{eq14}
\end{eqnarray}
\noindent
Eqs.~(\ref{eq4a}), (\ref{eq13}), and (\ref{eq14}) allow us to
place a lower limit on the magnetic field in the blob
\begin{eqnarray}
B > 15.1 t_{\rm var}^{-2/3} \epsilon_{b}^{-1/3} D^{-1/3}.
\label{eq15}
\end{eqnarray}

We next estimate the ICS cooling time and require this to be less
than the variability time scale.  For the soft photon spectrum we
adopt, some interactions with energetic electrons will be in the
Thomson regime, and others will be in the Klein-Nishina regime with
relatively small energy loss.  To calculate the cooling time, we
neglect interactions in the Klein-Nishina regime, i.e. with
photons above $\epsilon'_T \approx m_e c^2/\gamma'$, and use the
simple Thomson energy loss formula
\begin{eqnarray}
{t'}_{\rm cool}^{\rm ICS} = {{m_e c^2}\over{k U_{rad}(<\epsilon'_T) \gamma'}},
\label{eq16}
\end{eqnarray} 
\noindent
where
\begin{eqnarray}
U_{rad}(<\epsilon'_T) \approx \int_0^{\epsilon'_T} n(\epsilon') \epsilon'
d\epsilon' \approx 5 C \epsilon_T^{'0.2},
\label{eq17}
\end{eqnarray}
$C = 4 d^2/(c^3 t_{\rm var}^2 D^{4 + \beta_1})$,
and we have used the fact that $\epsilon'_T < \epsilon'_{b}$
for electrons emitting at TeV energies.

Since the recent report on the Mrk 421 flare observations by the
Whipple Observatory gives no evidence of a spectral break at high
energies, we assume that the break in the $\gamma$-ray spectrum
is below but close to the threshold of the Whipple observations,
i.e. at $\sim 0.3$ TeV. This in turn means that electrons with
Lorentz factor $\gamma'_{b}$ must cool during the $\gamma$-ray
flare, and so we shall use $\gamma'=\gamma'_b$ in Eq.~(\ref{eq16}) to
estimate the ICS cooling time scale.  Then, from
Eqs.~(\ref{eq13}) and (\ref{eq16}) we can obtain an upper limit
on the magnetic field in the blob as a function of Doppler
factor, as required by the ICS cooling argument
\begin{eqnarray}
B < 4.4\times 10^{31}\epsilon_{b}^2 t_{\rm var}^{-2.5} D^{-13}.
\label{eq19}
\end{eqnarray}
   \begin{figure}
      \vspace{14.cm}
\includegraphics{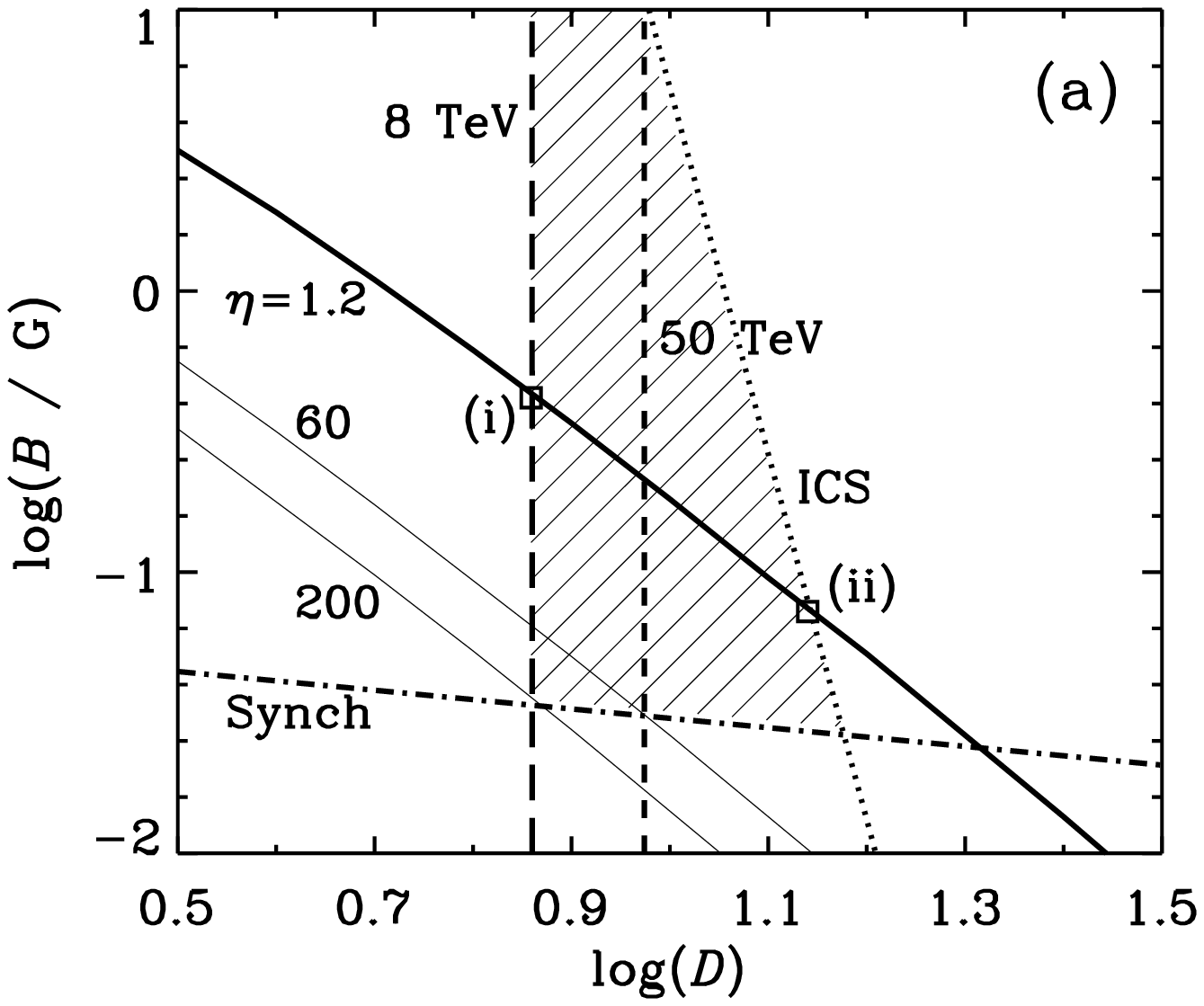}
\includegraphics{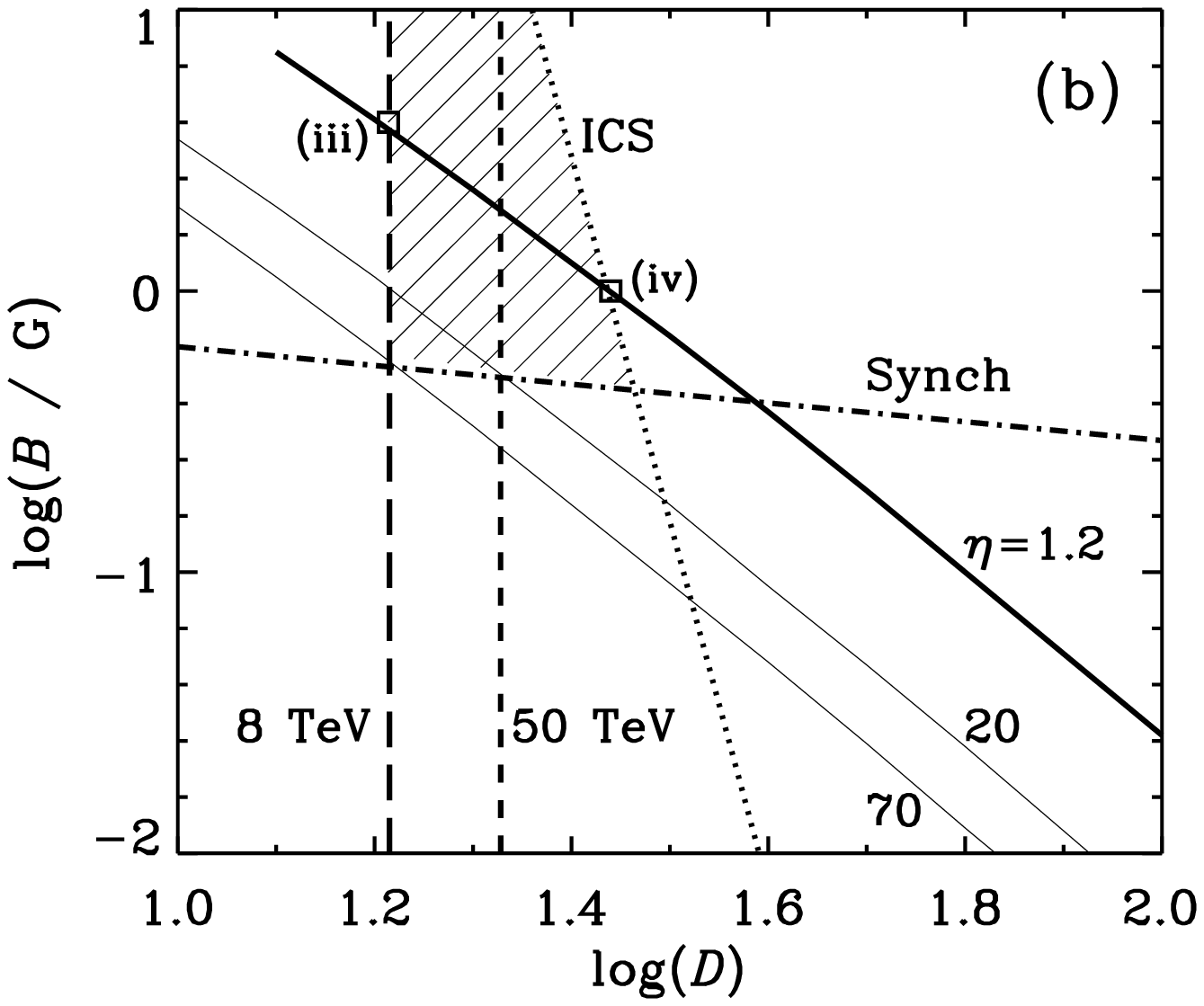}
      \caption[]{

The parameter space ($B$, $D$) allowed by the homogeneous SSC
model for variability in Mrk 421 on a time scale $t_{\rm var}$
of (a) 1~day and (b) 15~min. The thick full curves show the
condition for $\eta=1.2$, and the thin full curves (labelled by
value of $\eta$) show the condition for other values of $\eta$.
The other curves give allowed ranges for: efficient cooling of
electrons during the flare time scale by synchrotron radiation
(Eq.~\ref{eq15}) -- area above dot-dash curve labelled `Synch';
efficient cooling by ICS (Eq.~\ref{eq19}) -- area below dotted
curve labelled `ICS'; escape of 8 TeV $\gamma$-rays -- area to
right of long-dashed curve; escape of 50 TeV $\gamma$-rays --
area to right of short-dashed curve. The shaded area is the
allowed region for the parameters for a spectrum extending to
8~TeV. The marginal values of ($B$, $D$) which just fulfil the
condition $\eta = 1.2$ are marked by squares and labelled (i) to
(iv).}
\label{fig1}
    \end{figure}

The constraints on $B$ and $D$ of the blob, derived above
(Eqs.~\ref{eq15}, and~\ref{eq19}), are shown in
Figs.~\ref{fig1}(a) and \ref{fig1}(b) for the two variability
time scales.  In the next section, we derive an additional limit
on the Doppler factor of the blob which can be obtained based on
the non-observation of absorption of $\gamma$-rays by
photon-photon pair production with soft photons in the blob
radiation.

\subsection{Absorption of gamma-rays in the blob radiation}

The observation of $\gamma$-ray flares with a spectrum extending
up to $\sim 8$ TeV, or even 50 TeV, allows us to place a lower
limit on the Doppler factor of the blob under the assumptions of
the homogeneous SSC model.  Using the observed soft photon
spectrum of Mrk 421 (Eq.~\ref{eq3}) we can compute the optical
depth $\tau(E'_\gamma, D)$ for $\gamma$-ray photons with energy
$E'_\gamma$ for $e^\pm$ pair production inside the blob,
\begin{equation}
\tau(E'_\gamma, D) = {r_\gamma \over 8 {E'_\gamma}^2}
\int_{\epsilon'_{\rm min}}^{\infty} \, d\epsilon'
\frac{n(\epsilon')} {\epsilon'^2} \int_{s_{\rm min}}^{s_{\rm
max}(\epsilon',E'_\gamma)} ds \, s \sigma(s),
\label{eq:mpl}
\end{equation}
where $n(\epsilon')$ is the differential photon number density
and $\sigma(s)$ is the total cross section for photon-photon pair
production (Jauch and Rohrlich 1955) for a centre of momentum
frame energy squared given by
\begin{equation}
s=2 \epsilon' E'_\gamma(1 - \cos \theta)
\label{eq:s}
\end{equation}
where $\theta$ is the angle between the directions of the
energetic photon and soft photon, and
\begin{eqnarray}
s_{\rm min} &=& (2 m_e c^2)^2,\\
\epsilon_{\rm min} &=& {(2 m_e c^2)^2 \over 4E'_\gamma},\\
s_{\rm max}(\epsilon',E'_\gamma) &=& 4\epsilon' E'_\gamma.
\end{eqnarray}

The condition
\begin{eqnarray}
\tau(E'_\gamma, D) < 1.,
\label{eq:ggee}
\end{eqnarray}
\noindent
gives us the limit on $D$. These lower limits are shown by the
dashed lines in Figs.~\ref{fig1}(a) and~\ref{fig1}(b) for variability
time scales $t_{\rm var} = 1$ day and 15 min, and for $E_\gamma =8$
TeV (long-dashed lines, Krennrich et al.~1997) and 50 TeV
(short-dashed lines, Meyer \& Westerhoff~1996).  The
allowed regions of the parameter space ($B$, $D$) determined by
Eqs.  ~\ref{eq15},~\ref{eq19}, and~\ref{eq:ggee} are shown in
Figs.~\ref{fig1}(a) and \ref{fig1}(b) by the shaded areas.

\section{Discussion and Conclusion}

   \begin{figure}
      \vspace{7.2cm}
\includegraphics{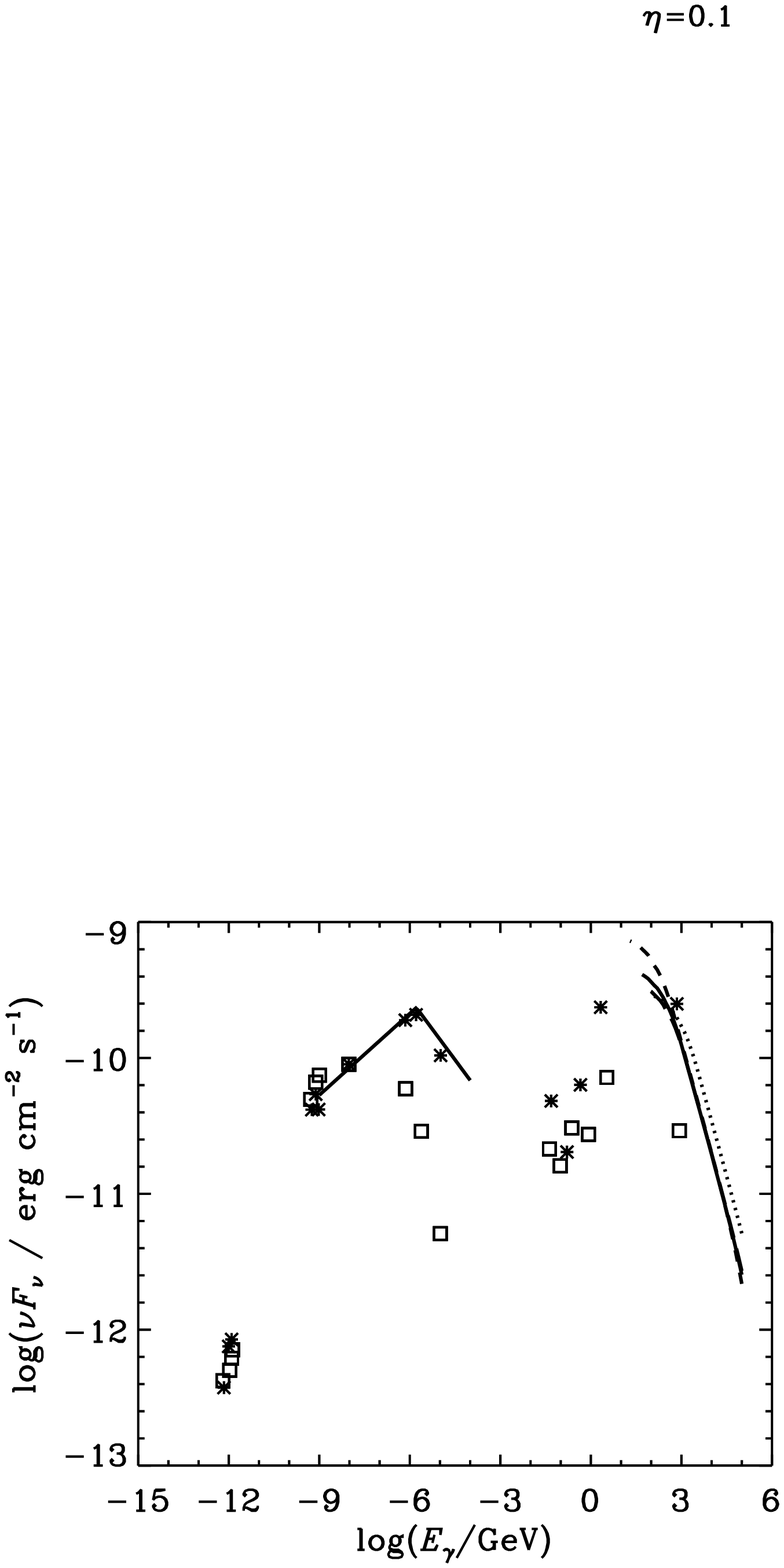}
      \caption[]{

The adopted soft photon spectrum (solid curve at $10^{-9}$ to
$10^{-4}$~GeV), and the ICS spectra calculated in terms of the
homogeneous SSC model for the four marginal values of ($B$, $D$)
of the allowed regions marked by the numbered open squares in
Figs.~\ref{fig1}(a) and~\ref{fig1}(b): (i) solid curve; (ii)
dotted curve; (iii) short-dashed curve; (iv) long-dashed curve.
Calculated spectra are compared with the observations of Mrk 421
(Macomb et al.~1995) during flaring (asterisk) and in a
quiscent state (open squares).}
\label{fig2}
    \end{figure}

Inspection of the Figs.~\ref{fig1}(a) and~\ref{fig1}(b) shows
that for some values of ($B$, $D$), i.e. the region of the thick
full line inside the shaded area, the homogeneous SSC model can
in principle produce flares with $\eta = 1.2$ as required.  We
note that the values of ($B$, $D$) used in earlier modeling of
the Mrk 421 spectrum (Inoue \& Takahara~1996, Mastichiadis \&
Kirk~1997, Stecker, De Jager \& Salamon~1996) are generally
consistent with the parameter space derived by us. In order to
determine if the broad band spectrum expected in the homogeneous
SSC model is consistent with the $\gamma$-ray observations during
flaring, we compute the synchrotron and ICS spectra for four
example parameters ($B$, $D$) from the allowed region indicated
by points (i) to (iv) in Figs.~\ref{fig1}(a) and \ref{fig1}(b).
The calculated spectra are shown in Fig~\ref{fig2}.  Note that in
each case, the lowest energy we predict corresponds to $\gamma' =
0.032 \gamma'_b$ (see Eq.~\ref{eq4} and comments below) which
depends on the magnetic field.  For the 4 cases this gives
$E'_\gamma \approx 36$~GeV (i), 120~GeV (ii), 19~GeV (iii), and
47~GeV (iv), for the minimum energies for which we can predict
the the $\gamma$-ray spectrum with any confidence in the
homogeneous SSC model.

For energies between 0.8 TeV and 8 TeV corresponding to
observations made by the Whipple observatory during recent
flaring of Mrk 421 (Krennrich et al.~1997), our predictions of
the spectral index in the homogeneous SSC model range from
2.65 for 1 day variability and case (i), to 2.85 for 15 minute
variability and case (iv).
The results obtained during flaring by Krennrich et al.~(1997) up
to $\sim 8$ TeV are consistent with the spectrum of Mohanty et
al.~(1993) taken during a quiescent state where the spectral
index was $\sim 2.25\pm 0.19\pm 0.3$ between $0.4 - 4$ TeV.
Given the error bars, this is just consistent with the spectral
index of 2.65 predicted for 1 day variability and case (i).
However, we note that the calculated spectrum shows
a break close to $\sim 1$ TeV which should be seen in the Whipple
observations.  

In the case of a flare varying on a 15 min time scale, it seems
that the spectra obtained in terms of the homogeneous SSC model
are not consistent with the relatively flat spectrum of the
Whipple observations.  The lower sensitivity HEGRA Cherenkov
observations report a very steep spectrum above $\sim 1$ TeV
(spectral index $3.6\pm 1.$) during the Dec. 94 - May 95
monitoring (Petry et al.~1996). However these observations refer
not to outburst emission, but rather to quiescent emission since
the spectrum is integrated over a long period.

In conclusion, detailed spectral measurements in the energy range
above 0.3 TeV combined with the observations in the
optical--X-ray range should allow one to determine precisely the
parameters of the emission region (relativistic blob) and in
general answer the question of the applicability of the
homogeneous SSC model for $\gamma$-ray production in blazars.  We
note also that the absorption and synchrotron cooling conditions
do not allow flares with 1 day variability having $\eta > 200$ (8
TeV) or $\eta > 60$ (50 TeV) -- see the thin solid curves in
Fig.~\ref{fig1}(a).  Similarly, for 15 minute variability $\eta >
70$ (8 TeV) or $\eta > 20$ (50 TeV) are not allowed (see
Fig.~\ref{fig1}b).  Observation of such huge $\gamma$-ray
outbursts without accompanying X-ray outbursts would be
inconsistent with the homogeneous SSC model.

\section*{Acknowledgements}

W.B. thanks the Department of Physics and Mathematical Physics at
the University of Adelaide for hospitality during his visit. This 
research is supported by a grant from the Australian Research Council.

\end{document}